\documentclass[rnote,tradiabstract]{aa} 
\usepackage{color}
\usepackage{graphicx}
\usepackage{natbib}
\usepackage[varg]{txfonts}
\usepackage{longtable,lscape}
\usepackage{url}
\usepackage{verbatim}
\usepackage{dcolumn}

\usepackage{txfonts}
\newcolumntype{d}[1]{D{.}{\cdot}{#1}}

\newcolumntype{.}{D{.}{.}{-1}}

\newcommand{\msun}{M$_\odot$}

\newcommand{\mum}{$\mu$m}

\newcommand{\sex}{\texttt{SExtractor}}
\newcommand{\gc}{\texttt{GaussClump}}

\newcommand{\rms}{r.m.s.}

\newcommand{\KS}{Kolmogorov-Smirnov}

\begin{document}
\bibliographystyle{aa-package/bibtex/aa}

\title{ATLASGAL --- Complete compact source catalogue: 280\degr\ $ <\ell <$ 60\degr}
\authorrunning{Urquhart et al.}
\titlerunning{The complete compact source catalogue}

\author { J.\,S.\,Urquhart\inst{1}\thanks{E-mail:
jurquhart@mpifr-bonn.mpg.de (MPIfR)}, T.\,Csengeri\inst{1}, F.\,Wyrowski\inst{1}, F.\,Schuller\inst{1,2}, S.\,Bontemps\inst{3}, L.\,Bronfman\inst{4}, K.\,M.\,Menten\inst{1}, C.\,M.\,Walmsley\inst{5,6}, Y.\,Contreras\inst{7},  H.\,Beuther\inst{8}, M.\,Wienen\inst{1} and H.\,Linz\inst{8}
  }

  \institute {Max-Planck-Institut f\"ur Radioastronomie, Auf dem H\"ugel 69,
    Bonn, Germany
    \and European Southern Observatory, Alonso de Cordova 3107, Vitacura, Santiago, Chile
     \and Laboratoire d'Astrophysique de Bordeaux –- UMR 5804, CNRS – Universit\'e Bordeaux 1, BP 89, 33270 Floirac, France 
    \and Departamento de Astronom\'{i}a, Universidad de Chile, Casilla 36-D, Santiago, Chile
    \and Osservatorio Astrofisico di Arcetri, Largo E. Fermi, 5, 50125 Firenze, Italy
    \and Dublin Institute for Advanced Studies,  31 Fitzwilliam Place, Dublin 2, Ireland
    \and CSIRO Astronomy and Space Science, PO Box 76, Epping, NSW 1710, Australia
    \and Max-Planck-Institut f\"ur Astronomie, K\"onigstuhl 17, 69117 Heidelberg, Germany
}

\date{Received xxx; accepted xxx}


\abstract
{The APEX Telescope Large Area Survey of the Galaxy (ATLASGAL) is the
  largest and most sensitive systematic survey of the inner Galactic plane in the
  submillimetre wavelength regime. The observations were carried out with the Large APEX
  Bolometer Camera (LABOCA), an array of 295 bolometers
  observing at 870\,$\mu$m (345 GHz).} {In this research note we present the compact source catalogue for the 280\degr\ $ <\ell <$ 330\degr\ and  21\degr\ $ <\ell <$ 60\degr\ regions of this survey.}{The construction of this catalogue was made with the source
 extraction routine \sex\ using the same input parameters and procedures used to analyse the inner Galaxy region presented in an earlier publication (i.e., 330\degr\ $ <\ell <$ 21\degr).}{We have identified 3524 compact sources and present a catalogue of their properties. When combined with the regions already published this provides a comprehensive and unbiased database of $\sim$10164 massive, dense clumps located across the inner Galaxy.}{} 
\maketitle
%

\section{Introduction}

The APEX Telescope Large Area Survey of the Galaxy (ATLASGAL) is an unbiased 870\,\mum\ submillimetre survey covering 420\,sq.\,degrees of the inner Galactic plane (\citealt{schuller2009}). The regions covered by this survey are $|\ell| < 60\degr$ with $|b|< 1.5\degr$ and $280\degr < \ell < 300\degr$ with $b$ between $-$2\degr\ and 1\degr; the change in latitude was necessary to take account of the warp in the Galactic disc in the outer Galaxy extension. The thermal emission from dust is optically thin at submillimetre wavelengths and contribution from free-free emission should be negligible making it an excellent tracer of column density and total clump mass. The fundamental goal of ATLASGAL is to provide a large and systematic inventory of massive, dense clumps in the Galaxy that includes representative samples of all of the earliest embedded stages of high-mass star formation.

\begin{figure}
\begin{center}
\includegraphics[width=0.49\textwidth, trim= 0 0 0 0]{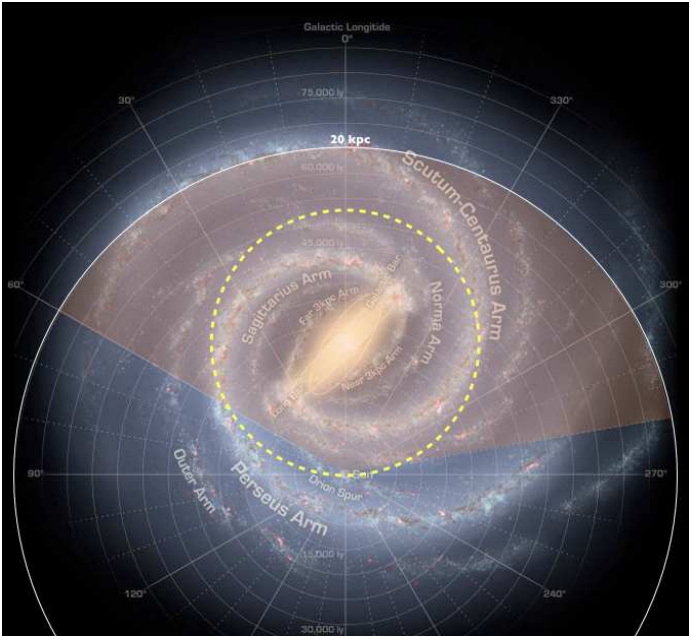}

\caption{\label{fig:coverage} Schematic of the Galaxy showing the region covered by ATLASGAL. The background image is an artist impression of the Galactic plane and incorporates much of what is known about the large-scale structure of the Galaxy and shows the location of the spiral arms and the Galactic bar. The Solar Circle is shown in by the dashed yellow line and the region of the Galaxy covered by the ATLASGAL survey shown by the light brown shading. Background image courtesy of NASA/JPL-Caltech/R. Hurt (SSC/Caltech).}
\end{center}
\end{figure}

ATLASGAL is the largest and most sensitive submillimetre ground-based survey and covers approximately 2/3 of the surface area of the Galactic molecular disc within $\sim$15\,kpc of the Galactic Centre (see Fig.\,\ref{fig:coverage} for a schematic of survey coverage). This increases to almost 100\% of the surface area of the molecular disc within the Solar circle (i.e., $R_{\rm{gc}} < 8.5$\,kpc) and includes all of the 5\,kpc molecular ring where the majority of the molecular gas in the Galaxy resides (i.e., $4 < R_{\rm{gc}} < 6$\,kpc; \citealt{stecker1975}). This survey has a sensitivity of 0.3-0.5\,Jy\,beam$^{-1}$ (5$\sigma$; see Sect.\,2 for details) and is able to detect all cold dense clumps ($<25$\,K) with masses $\ge$1000\,\msun\ at a heliocentric distance of 20\,kpc, which is sufficient to find all massive star forming clumps located in the inner Galaxy (assuming a standard initial mass function and a typical star formation efficiency; \citealt{urquhart2013a}). Furthermore, this survey complements many of the other Galactic plane surveys (e.g., UKIDSS, \citealt{lucas2008});  GLIMPSE, \citealt{benjamin2003_ori}; \textit{Herschel} Hi-GAL, \citealt{molinari2010}); Bolocam Galactic Plane Suvey, \citealt{aguirre2011}  ; and the Co-ordinated Radio and Infrared Survey for High-Mass Star Formation (CORNISH), \citealt{purcell2013} and \citealt{hoare2012}).

To date, two methods have been used to identify sources present in the ATLASGAL emission maps and extract their parameters. The first method used was the \sex\ extraction algorithm (\citealt{bertin1996}) that was applied to a $\sim$150 sq. degree region located towards the Galactic Centre (i.e., $330\degr < \ell < 21\degr$; \citealt{contreras2013}). This method is sensitive to the large-scale structures of dense clumps and small molecular clouds and identified 6640 sources above a threshold of 3$\sigma$ ($\sim$180-300\,mJy\,beam$^{-1}$) and is 99\% complete above 6$\sigma$. The second method consists of a two step process where the emission maps are first filtered to remove the large-scale diffuse structure, and then the \gc\ algorithm is used to identify and extract source parameters by fitting a 2D Gaussian to the structures. This method has been applied to the whole ATLASGAL survey region (i.e., $280\degr < \ell < 60\degr$) and has identified almost 11000 compact embedded objects above a more conservative 5$\sigma$ threshold (\citealt{csengeri2014}). 

We refer to the \sex\ and \gc\ catalogues as the Compact Source Catalogue (CSC) and GaussClump Source Catalogue (GCSC), respectively.\footnote{\texttt{http://atlasgal.mpifr-bonn.mpg.de/cgi-bin/ATLASGAL\_DATABASE.cgi}} The two methods have been designed  to be sensitive to different structures with the former providing information on the global parameters of the whole clump, while the latter is better to estimate the properties of the gas more closely associated with the star formation process (see Sect.\,3 for more details).

In this research note we present the results of the \sex\ source extraction for the region not covered by \citet{contreras2013}. A detailed description of the survey is provided by \citet{schuller2009} and a comprehensive discussion of the \sex\ algorithm, procedures used, catalogue completeness and reliability tests can be found in \citet{contreras2013}. We provide only a brief summary of the ATLASGAL survey and the source extraction method in Sect.\,\ref{sect:survey_details} and refer the reader to the these two papers for more details. In Sect.\,\ref{sect:results} we present the catalogue of new sources and compare their properties with those identified by \citet{contreras2013} and with the GCSC (\citealt{csengeri2014}). We present a summary and outlook in Sect.\,\ref{sect:summary}.

\section{Survey and source extraction summary}
\label{sect:survey_details}

The observations were carried out with the APEX
(Atacama Pathfinder EXperiment) 12\,m submillimetre telescope \citep{gusten2006},
located in Llano de Chajnantor, Chile. The Large APEX BOlometer CAmera (LABOCA; \citealt{siringo2009}) was used to map the 870\,\mum\ emission across the inner Galaxy. The instrument bandpass is centred at 345\,GHz and has a bandwidth of 60\,GHz, which corresponds to FWHM beam size of $19{\rlap.{}''}2$. The typical pointing \rms\ error is $\sim$\,4$''$. The noise varies between 60 and 100\,mJy\,beam$^{-1}$ across the survey region depending on the number of coverages, the atmospheric conditions at the time of the observations and telescope elevation. The lower values are typically found towards the Galactic Centre regions and the highest values are associated with the extension region (i.e., $280\degr < \ell < 300\degr$; see Fig.\,1 of \citealt{csengeri2014}). 

\sex\ was used to identify coherent emission in the signal noise ratio (SNR) maps above a  threshold value of 3 and associated with more pixels than the beam integral (i.e., $\sim$11 pixels). We have used the same input parameters and emission maps reduced in the same way as those used by \citet{contreras2013} and therefore represents a natural extension to their work and combine provides a fully consistent catalogue covering the whole ATLASGAL survey. The SNR maps were produced by dividing the emission maps by the square root of the weight maps; using these avoids problems encountered when trying to identify sources using a flux threshold in fields where the background noise is varying. The parameters of the sources above the threshold are then extracted from the original emission maps. 

Tests were performed to check the reliability of the \sex\ algorithm and the catalogue of extracted sources. This was done by injecting artificial sources into two types of synthetic maps; one where the noise properties were Gaussian and another where the noise included a varying component introduced to better reflect the diffuse emission seen in the survey emission maps. \sex\ recovered 99\% of injected sources in both types of maps above a threshold of $\sim$6$\sigma$ and this is taken as the completeness of the catalogue and  for reliable detections (for more details see Sect.\,4 of \citealt{contreras2013}).

\begin{figure*}
\begin{center}
\includegraphics[width=0.45\textwidth, trim= 0 0 0 0]{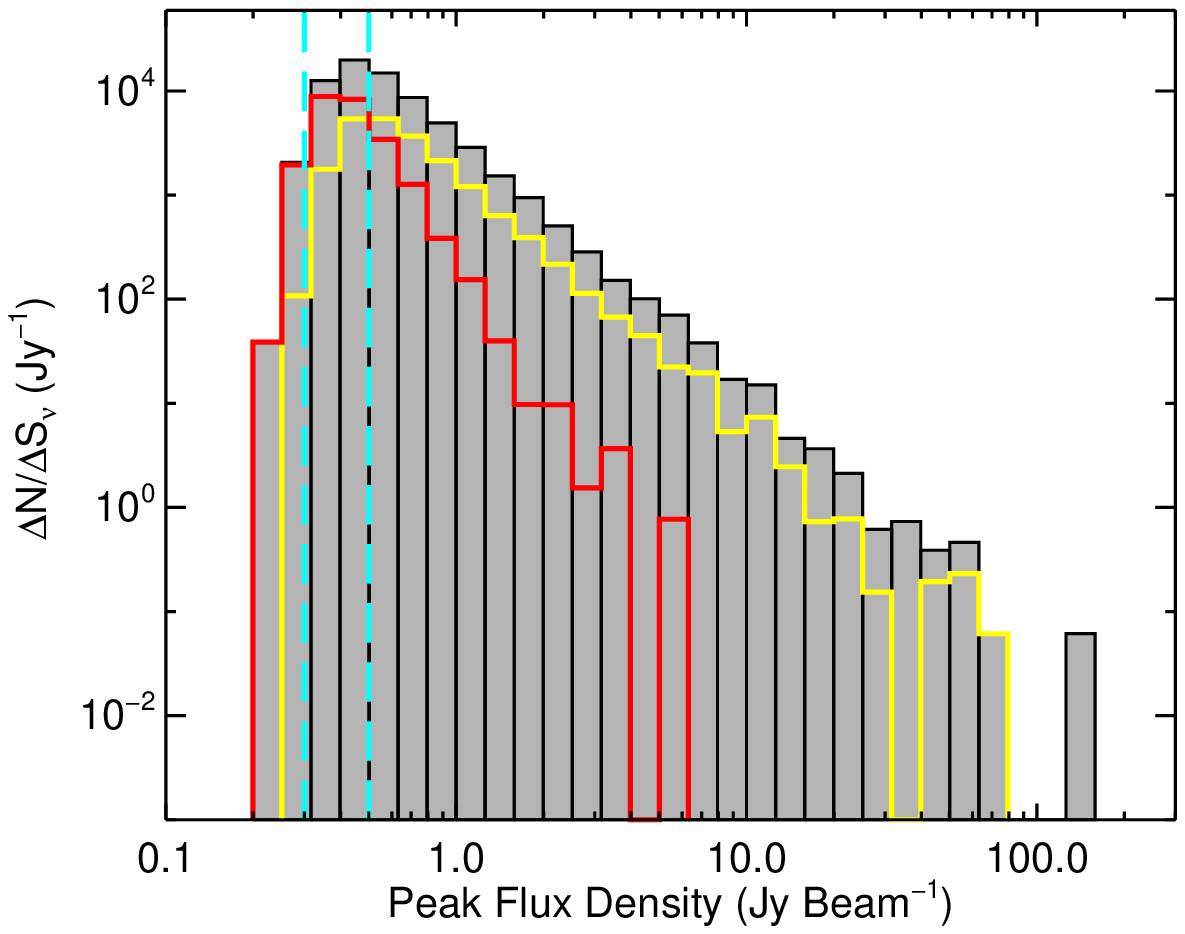}
\includegraphics[width=0.45\textwidth, trim= 0 0 0 0]{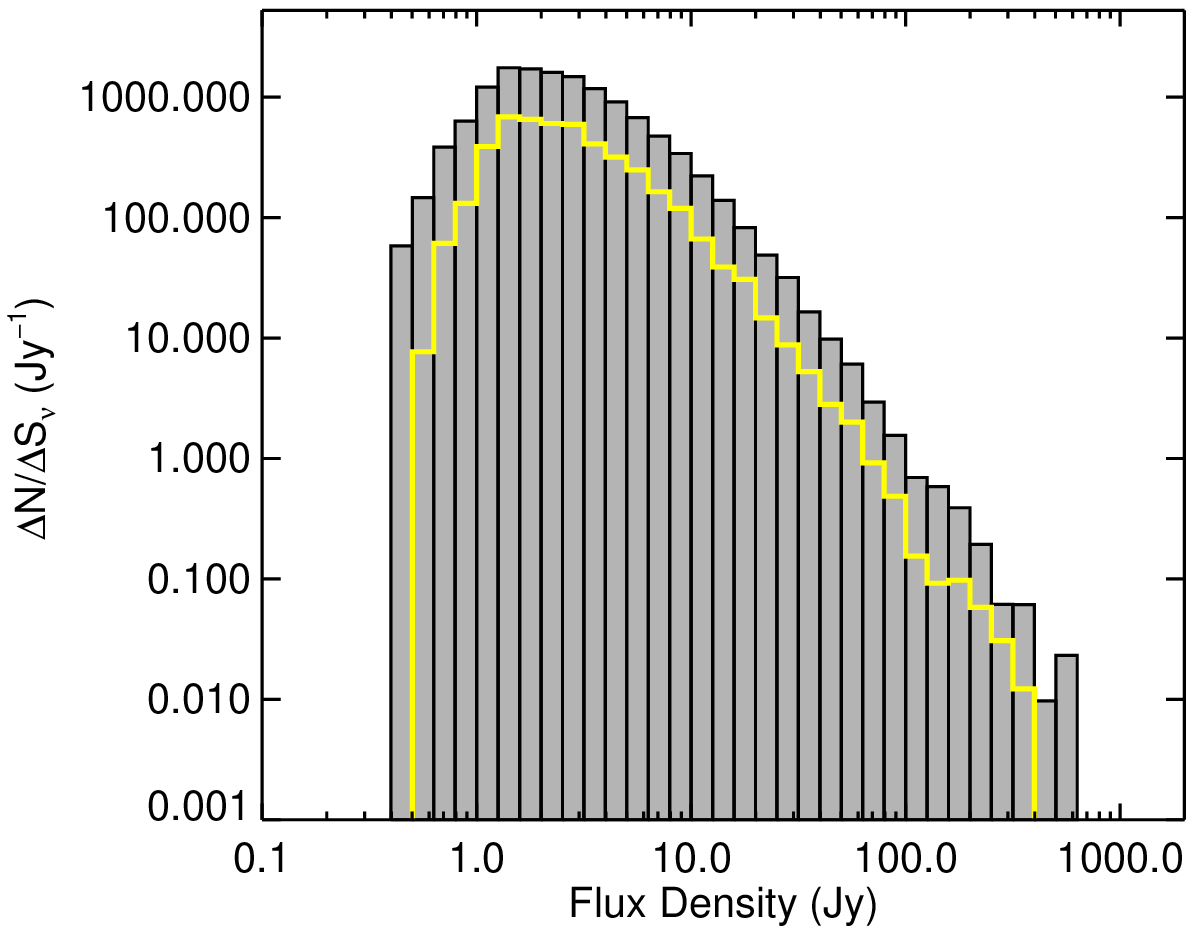}

\caption{\label{fig:flux_density} Flux density distribution for the full ATLASGAL CSC extracted by \sex\ (grey filled histogram) and the sources located in the 280\degr\ $ <\ell <$ 330\degr\ and  21\degr\ $ <\ell <$ 60\degr\ region (yellow). In the left and right panels we present histograms of the peak and integrated flux densities measured for each source respectively. The dashed vertical lines shown in the left panel indicate the nominal 6$\sigma$ sensitivity range above which the catalogue is considered complete (i.e., 0.3-0.5\,Jy\,beam$^{-1}$). The red histogram shows the peak flux distribution of sources not matched with a source identified by the GaussClump catalogue. The bin width is 0.1\,dex.}

\end{center}
\end{figure*}

\section{Compact source catalogue}
\label{sect:results}
\subsection{Catalogue description}

We have run \sex\ on thirty $3\degr\times 3\degr$ fields outside the region previously reported by \citealt{contreras2013} (i.e., 280\degr\ $ <\ell <$ 330\degr\ and  21\degr\ $ <\ell <$ 60\degr) and identified a further 3524 compact sources in this 267\,sq.\,degree region. In Table\,\ref{tbl:cattable} we present a sample of the catalogue as an
example of its form and content. In this table we present the source name, the positions of the emission peak and its geometric centre, the size of the semi-major and minor axis and their position angle, the deconvolved radius, the peak and integrated fluxes and their associated errors and any warning flags generated by the algorithm identify possible problems or artefacts affecting the source. 

The deconvolved source sizes are estimated assuming Gaussian beam characteristics following \citet{rosolowsky2010} such that: 

\begin{equation}
\label{radius}
\theta_{\rm{R}}= \eta \left[(\sigma_{\rm{maj}}^2-\sigma_{\rm{bm}}^2)(\sigma_{\rm{min}}^2-\sigma_{\rm{bm}}^2)\right]^{1/4},
\end{equation}

\noindent where $\sigma_{bm}$ is the 1$\sigma$ Gaussian beam size (i.e.,
$\sigma_{bm}=\theta_{\mathrm{FWHM}}/\sqrt{8\ln 2} \simeq 8''$) and a value for 
$\eta$ of 2.4 \citep{rosolowsky2010} that relates the measured size to the source's effective radius (i.e., $R_{\rm{eff}}=\sqrt(A/\pi)$, where $A$ is the surface area of the source; \citealt{dunham2011}). An important caveat is that since the source sizes are determined only using the pixels above the detection threshold we are likely to underestimate the true source sizes, and in some cases this can result in source sizes that are smaller than the beam. 

The errors given for the peak and integrated fluxes include the absolute calibration uncertainty of 15\% and the intrinsic measurement error, which are added in quadrature. In Col.\,14 we give the quality flag provided by \sex, which identifies possible problems or artefacts affecting the source  (see \citealt{bertin1996} for details) and in the final column we give the signal to noise ratio for each source.

\subsection{Catalogue properties}

Combined with the sources previously identified from analysis of the $330\degr < \ell < 21\degr$ region reported by \citet{contreras2013} the complete ATLASGAL CSC consists of 10170 sources. In Fig.\,\ref{fig:flux_density} we present plots showing the peak and integrated flux densities for the full CSC catalogue (grey filled histogram) and for comparison the new sources reported here (yellow histogram). The slopes of the two distributions are very similar and a power-law fit to the bins above the turnover ($\sim$0.5\,Jy\,beam$^{-1}$; corresponds to the nominal completeness) finds slopes of $-2.43\pm0.04$ and $-2.41\pm0.07$ for the full catalogue and new sources, respectively. The peak flux distributions are therefore in excellent agreement with each other and the value reported by \citet{contreras2013} for the inner Galaxy region (i.e., $-2.4\pm0.04$). Analysis of the integrated flux distribution finds similar agreement between the full CSC and the sources reported here and by \citet{contreras2013}. We therefore conclude that there are no systematic differences in the flux properties of the whole \sex\ catalogue.

\begin{figure}
\begin{center}
\includegraphics[width=0.45\textwidth, trim= 0 0 0 0]{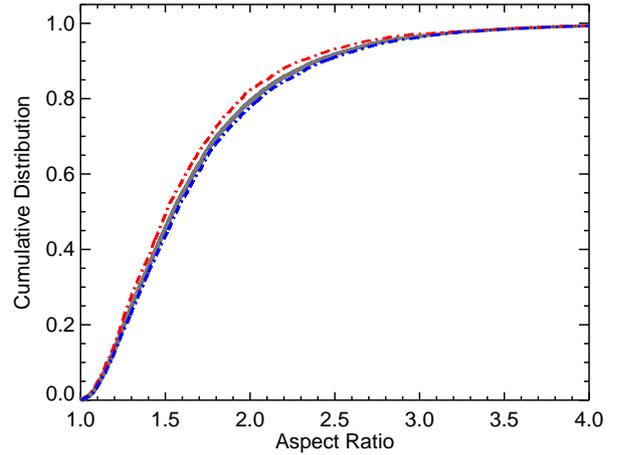}

\caption{\label{fig:aspect_ratio} The cumulative distributions of the aspect ratio for the full ATLASGAL CSC (grey curve) and the sources reported here and by \citet{contreras2013} are shown by the red and blue curves, respectively.}

\end{center}
\end{figure}

In Fig.\,\ref{fig:aspect_ratio} we present a cumulative distribution plot showing the measured aspect ratio of the full catalogue (i.e., $\theta_{\rm{maj}}$/$\theta_{\rm{min}}$) and the \citet{contreras2013} subsample and the new sources reported here. Inspection of these distributions reveals the new sources have slightly smaller aspect ratios than the \citet{contreras2013} subsample (a \KS\ test finds this difference is significant with a probability  $\ll 0.01$ that the two samples are drawn from the same parent distribution, i.e., less than 3$\sigma$). However, the source density in the 280\degr\ $ <\ell <$ 330\degr\ and  21\degr\ $ <\ell <$ 60\degr\ regions is significantly lower and the overall emission is less complicated and there is less blending of sources, which is likely to result in more elongated sources being found towards the innermost part of the Galactic plane.

\setcounter{table}{0}

\setlength{\tabcolsep}{4pt}

\begin{table*}

\begin{center}\caption{\label{tbl:cattable} The ATLASGAL compact source catalogue for sources located in the 280\degr\ $ <\ell <$ 330\degr\ and  21\degr\ $ <\ell <$ 60\degr\ region.  The columns are as follows: (1) name derived from Galactic coordinates of the maximum intensity in the source; (2)-(3) Galactic coordinates of maximum intensity in the catalogue source; (4)-(5) Galactic coordinates of emission centroid; (6)-(8) semi-major and semi-minor size and source position angle measured anti-clockwise from Galactic north; (9) effective radius of source; (10)-(13) peak and integrated flux densities and their associated uncertainties; (14) \sex\ detection flag (see \citealt{bertin1996} notes on these flags); (15) signal to noise ratio (SNR) --- values for sources with peak flux below $6\sigma$ detection should not be used blindly.}
\begin{minipage}{\linewidth}
\begin{tabular}{l....rrrr....c.}
  \hline \hline
  \multicolumn{1}{c}{Catalogue} 
  &  \multicolumn{1}{c}{$\ell_{\mathrm{max}}$} &  \multicolumn{1}{c}{$b_{\mathrm{max}}$}
  &  \multicolumn{1}{c}{$\ell$} &  \multicolumn{1}{c}{$b$} &
  \multicolumn{1}{c}{$\sigma_{\rm{maj}}$} &  \multicolumn{1}{c}{$\sigma_{\rm{min}}$} &  \multicolumn{1}{c}{PA} &
  \multicolumn{1}{c}{$\theta_{\rm{R}}$} &  \multicolumn{1}{c}{$S_{\rm{peak}}$} & \multicolumn{1}{c}{$\Delta S_{\rm{peak}}$} & \multicolumn{1}{c}{$S_{\rm{int}}$}& \multicolumn{1}{c}{$\Delta S_{\rm{int}}$} & \multicolumn{1}{c}{Flag} & \multicolumn{1}{c}{SNR} \\
   
  \multicolumn{1}{c}{Name} &  \multicolumn{1}{c}{($^{\circ}$)} &
  \multicolumn{1}{c}{($^{\circ}$)} &  \multicolumn{1}{c}{($^{\circ}$)} &
  \multicolumn{1}{c}{($^{\circ}$)} &  \multicolumn{1}{c}{($''$)}
  &  \multicolumn{1}{c}{($''$)} &  \multicolumn{1}{c}{($^{\circ}$)}&  \multicolumn{1}{c}{($''$)}
  &  \multicolumn{2}{c}{(Jy\,beam$^{-1}$)} &  \multicolumn{2}{c}{(Jy)}&\\
 
  \multicolumn{1}{c}{(1)} &  \multicolumn{1}{c}{(2)} &  \multicolumn{1}{c}{(3)} &  \multicolumn{1}{c}{(4)} &
  \multicolumn{1}{c}{(5)} &  \multicolumn{1}{c}{(6)} &  \multicolumn{1}{c}{(7)} &  \multicolumn{1}{c}{(8)} &
  \multicolumn{1}{c}{(9)} &  \multicolumn{1}{c}{(10)} &  \multicolumn{1}{c}{(11)} & \multicolumn{1}{c}{(12)} &  \multicolumn{1}{c}{(13)} &  \multicolumn{1}{c}{(14)} &  \multicolumn{1}{c}{(15)} \\
  \hline
AGAL300.164$-$00.087	&	300.164	&	-0.087	&	300.164	&	-0.089	&	11	&	7	&	10	&	$\cdots$	&	1.39	&	0.26	&	5.09	&	1.02	&	0	&	8.9	\\
AGAL300.218$-$00.111	&	300.218	&	-0.111	&	300.218	&	-0.113	&	19	&	12	&	122	&	30	&	1.17	&	0.26	&	5.68	&	1.11	&	0	&	6.2	\\
AGAL300.323$-$00.199	&	300.323	&	-0.199	&	300.323	&	-0.199	&	8	&	6	&	138	&	$\cdots$	&	0.71	&	0.38	&	2.28	&	0.57	&	16	&	1.9	\\
AGAL300.341$-$00.214	&	300.341	&	-0.214	&	300.340	&	-0.213	&	13	&	9	&	140	&	16	&	0.76	&	0.30	&	5.30	&	1.05	&	2	&	2.7	\\
AGAL300.381$-$00.287	&	300.381	&	-0.287	&	300.379	&	-0.286	&	25	&	13	&	141	&	37	&	0.90	&	0.14	&	7.11	&	1.33	&	0	&	97.4	\\
AGAL300.403+00.544	&	300.403	&	0.544	&	300.404	&	0.544	&	13	&	10	&	129	&	19	&	1.02	&	0.61	&	3.38	&	0.75	&	0	&	1.7	\\
AGAL300.456$-$00.189	&	300.456	&	-0.189	&	300.455	&	-0.191	&	14	&	9	&	15	&	17	&	0.94	&	0.17	&	4.29	&	0.89	&	0	&	9.9	\\
AGAL300.491$-$00.176	&	300.491	&	-0.176	&	300.489	&	-0.175	&	20	&	12	&	172	&	31	&	0.68	&	0.15	&	6.12	&	1.18	&	3	&	6.5	\\
AGAL300.499$-$00.212	&	300.499	&	-0.212	&	300.500	&	-0.211	&	17	&	7	&	7	&	$\cdots$	&	0.58	&	0.12	&	2.52	&	0.61	&	3	&	6.5	\\
  \hline
\end{tabular}\\
\end{minipage}
Notes: Only a small portion of the data is provided here, the full table is only  available in electronic form at the CDS via anonymous ftp to cdsarc.u-strasbg.fr (130.79.125.5) or via http://cdsweb.u-strasbg.fr/cgi-bin/qcat?J/A+A/.

\end{center}
\end{table*}

\setlength{\tabcolsep}{6pt}

\subsubsection{Comparison with the GaussClump catalogue}

\begin{figure}
\begin{center}
\includegraphics[width=0.49\textwidth, trim= 40 0 0 0, clip=true]{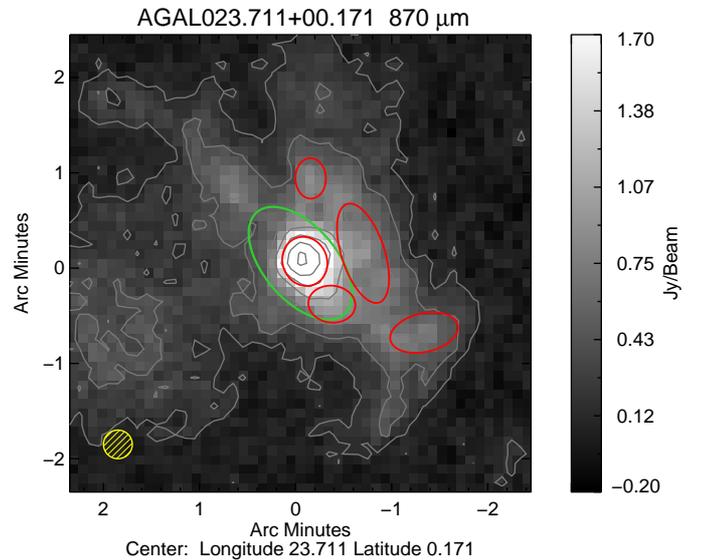}

\caption{\label{fig:sex_gc_example} Example of a source detected by both extraction algorithms but where \gc\ has separated the emission into multiple components. The greyscale shows the ATLASGAL dust emission and the red hatched circle shown in the lower left corner indicates the APEX FWHM beam size. The position and size of the CSC source is indicated by the green ellipse and the multiple GCSC sources are indicated by the yellow ellipses. Contours start and increase in steps of 3$\sigma$.}

\end{center}
\end{figure}

A comparison between the two catalogues is presented by \citet{csengeri2014} and so we present only a brief summary and discuss a few points not previously mentioned. Both catalogues contain a similar number of sources with 10164 identified by \sex\ and 10861 identified by GaussClumps (\citealt{csengeri2014}), however, only approximately 7600 CSC and 10400 GCSC sources are common to both catalogues, with $\sim$1600 CSC sources being associated with 2 or more GCSC sources (see Fig.\,\ref{fig:sex_gc_example} for an example). There are $\sim$500 GCSC sources without a counterpart in the CSC, however, in the majority of cases (90\%) a source was detected by \sex\ but these were later excluded as they had fewer than 11 pixels. There are some $\sim$2500 CSC sources without GCSC counterparts but the majority of these fall below the 5$\sigma$ threshold required for inclusion in that catalogue. In the left panel of Fig.\,2 we show the flux distribution of CSC sources without a GCSC counterpart (red histogram) that shows the unmatched sources dominate the lower peak flux density part of the distribution. The remaining unmatched CSC sources found are more extended regions of emission that were removed by the large-scale filtering in the GCSC catalogue.

The differences noted here nicely illustrate the complementary strengths of the two catalogues with the CSC providing a more global distribution of the dust while the GCSC has better sensitive to even more compact sources (as illustrated in Fig.\,4).

\section{Summary}
\label{sect:summary}

The ATLASGAL has mapped 420\,deg$^2$ of the inner Galaxy at
870\,$\mu$m, which traces thermal emission from dense clumps. In this research note we present analysis of $\sim$270 sq. degrees not included in the previous Compact Source Catalogue (CSC; \citealt{contreras2013}). This analysis has identified an additional 3524 compact sources located within the 280\degr\ $ <\ell <$ 330\degr\ and  21\degr\ $ <\ell <$ 60\degr\ regions of the Galactic disc. Combined with the catalogue presented by \citet{contreras2013} the full CSC now consists of 10164 compact sources and provides a complete inventory of massive star forming clumps with masses $\ge$1000\,\msun\ within the inner Galactic plane as well as large, but less complete, samples of lower mass clumps. This catalogue has already been used to identify large samples of massive star forming clumps and investigate their properties (i.e., \citealt{urquhart2013a}, \citealt{urquhart2013b} and Urquhart et al. 2014) and will be an indispensable resource for future studies with ALMA.$^1$

\begin{acknowledgements}
We would like to thank the referee Viktor T\'oth for his helpful comments and suggestions. This work was partially funded by the ERC Advanced Investigator Grant GLOSTAR (247078) and was partially carried out within the Collaborative Research Council 956, sub-project A6, funded by the Deutsche Forschungsgemeinschaft (DFG). L.B. acknowledges support from CONICYT through project BASAL PFB-06.
  
\end{acknowledgements}

\bibliography{rms}

\end{document}